\documentclass[letterpaper, 10 pt, conference]{ieeeconf}  % Comment this line out if you need a4paper

\IEEEoverridecommandlockouts                              % This command is only needed if 
\usepackage{graphics} % for pdf, bitmapped graphics files
\usepackage{epsfig} % for postscript graphics files
\usepackage{mathptmx} % assumes new font selection scheme installed
\usepackage{times} % assumes new font selection scheme installed
\usepackage{amsmath} % assumes amsmath package installed
\usepackage{amssymb}  % assumes amsmath package installed
\usepackage{amsmath,amssymb,wasysym,epsfig,color,subfigure,graphicx}
\usepackage{enumerate,url,algpseudocode,algorithm,balance,url,bm,nicefrac,caption,subfloat}

\usepackage[table]{xcolor}
\allowdisplaybreaks

\title{\LARGE \bf
A Statistical Learning Approach to Reactive 
Power \\
Control in Distribution Systems
}

\author{Qiuling Yang, Alireza Sadeghi, Gang Wang, 
	Georgios B. Giannakis, and Jian Sun% <-this % stops a space
\thanks{The work of Q. Yang and J. Sun was supported by the National Natural Science Foundation of China under Grants U1509215,  61621063, and 61720106011. 
Q. Yang was also supported by the China Scholarship Council. 
The work of A. Sadeghi, G. Wang, and G. B. Giannakis was supported by the NSF Grants 1509040, 1711471, and 1901134. % <-this % stops a space
Q. Yang and J. Sun are with the State Key Lab of Intelligent Control and Decision of Complex Systems, School of Automation, Beijing Institute of Technology, Beijing 100081, China.
A. Sadeghi, G. Wang, and G. B. Giannakis are with the Department of Electrical and Computer Engineering, University of Minnesota, Minneapolis, MN 55455, USA.
E-mail:  \{yang6726, sadeghi, gangwang, georgios\}@umn.edu; sunjian@bit.edu.cn.}
}

\begin{document}

\maketitle
\thispagestyle{empty}
\pagestyle{empty}

%%%%%%%%%%%%%%%%%%%%%%%%%%%%%%%%%%%%%%%%%%%%%%%%%%%%%%%%%%%%%%%%%%%%%%%%%%%%%%%%
\begin{abstract}

Pronounced variability due to the growth of renewable energy sources, flexible loads, and distributed generation is challenging residential distribution systems. This context, motivates well fast, efficient, and robust reactive power control. Real-time optimal reactive power control is possible in theory by solving a non-convex optimization problem based on the exact model of distribution flow. However, lack of high-precision instrumentation and reliable communications, as well as the heavy computational burden of non-convex optimization solvers render computing and implementing the optimal control challenging in practice. Taking a statistical learning viewpoint, the input-output relationship between each grid state and the corresponding optimal reactive power control is parameterized in the present work by a deep neural network, whose unknown weights are learned offline by minimizing the power loss over a number of historical and simulated training pairs. In the inference phase, one just feeds the real-time state vector into the learned neural network to obtain the `optimal' reactive power control with only several matrix-vector multiplications. The merits of this novel statistical learning approach are computational efficiency as well as robustness to random input perturbations. Numerical tests on a $47$-bus distribution network using real data corroborate these practical merits.

\end{abstract}

	\section{INTRODUCTION }\label{sec:intr}
Reliability and operational efficiency of modern distribution systems are currently being challenged by high penetration of unpredictable renewable energy resources, large-scale deployment of electric vehicles, and `human-in-the-loop' demand response programs. As a consequence, reverse power flow as well as voltage magnitude fluctuations are prevailing in nowadays residential grids \cite{carvalho2008distributed}. For instance, solar power generation may drop by $15 \%$ of the photo voltaic (PV) nameplate rating within one minute due, for example, to intermittent cloud coverage \cite{barker2012smart}, which will result in a sizable voltage sag if no action is taken. The role of networked control in power systems is to maintain desired operations, while preventing contingency
events involving voltage and/or frequency instabilities from developing into large-scale cascades and blackouts.
To protect electrical devices, bus voltage magnitudes in distribution grids are typically regulated to be within a certain range, e.g., $\pm 5\%$ around their nominal values. A common practice to achieve this is through reactive power compensation.	
%	To stabilize bus voltage magnitudes within the prescribed limits, e.g., $\pm 5\%$ of their nominal values and minimize power losses over distribution lines, reactive power management is practically appealing.   

Traditional approaches have relied on utility-owned devices including load-tap-changing transformers, voltage regulators, and capacitor banks to control reactive power injection into the grid. 
%	\cite{kersting2006distribution}. 
Although these devices perform well in certain cases, slow response times, discrete control actions, and lifespan limitations discourage them from fast reactive power control \cite{farivar2011inverter}. 
Recent advances in smart inverters offer new opportunities by circumventing these limitations. 
Despite their advantages, computing the optimal setpoints for smart inverters can be cast as an instance of the optimal power flow task, which entails solving a non-convex optimization problem \cite{farivar2011inverter,kekatos2015stochastic}. Furthermore, to deal with the renewable energy uncertainties as well as unreliable communication links (which cause delay and even communication failures),  
stochastic, online, decentralized, and localized smart inverter control schemes have been developed \cite{kekatos2015stochastic,wang2016ergodic,lin2018real,zhu2016fast,zhang2018distributed}. 
Nonetheless, centralized solvers suffer from high computational complexity, and decentralized and localized schemes 
%require high-precision instrumentation and reliable communications exchanges among inverters, and online and local control 
algorithms converge slowly.

To bypass these hurdles, recent proposals have engaged machine learning approaches for fast networked control and monitoring
%parameterization to optimize reactive power control
 \cite{oscar2019regression,yang2019real,tsp2019psse,zamzam2019psse, zamzam2019learning}.
% Practical advancements in machine learning and improvements in artificial intelligence can enabled the possibility of learning to optimize reactive power control in distribution systems. 
%	According to the learning mode, machine learning algorithms are divided into three categories: supervised learning, unsupervised learning, and reinforcement learning \cite{goodfellow2016deep, lecun2015deep}.
A support vector machine-based method was devised in \cite{kekatos2019designing} to approximate a near-optimal inverter control rule. In \cite{yang2019real}, the authors developed a
voltage regulation scheme using deep reinforcement learning.
% A batch reinforcement learning scheme using linear function approximation was developed for voltage regulation ~\cite{xu2018optimal}. 
% However, it is hard to design the reward function for the mentioned reinforcement learning algorithms.
%Evidenced by numerous empirical success, deep neural networks (deep neural networks) can find the optimal reactive power control rule without requiring any model for underlying distribution system \cite{aggarwal1998artificial},
% which is possible due to universal approximation property of deep neural networks \cite{cheng2019dnn}. 
%Due to the well known universal approximation property \cite{mark2019learning}, 
%Practical advancements in machine learning (ML) and improvements in artificial intelligence (AI), have further improved power system management. 
Deep (recurrent) neural networks were used for power system state estimation and forecasting in \cite{tsp2019psse}. By exploiting the power grid topology, a physics-aware neural network was proposed for state estimation \cite{zamzam2019psse}.
Related schemes leveraging deep neural networks that `learn-to-optimize' also appeared in resource allocation \cite{deeppower2018cl} and outage detection \cite{outage2019tsg}.
Unfortunately, training existing supervised learning models for reactive power control, requires large-scale labeled training data, which are difficult to be obtained in real-world physical systems. Reinforcement learning approaches on the other hand, entail prior knowledge on designing the so-called reward functions and often converge slowly.

Different from existing efforts, in this work an unsupervised statistical learning approach is developed for computationally intensive and time-sensitive reactive power control. 
%	This work combines machine learning tools with physical system, to advocate a deep neural network-based approach for designing reactive power control rules. % Our main contributions are on two fronts.
Specifically, a deep neural network is used to parameterize the functional relationship between the grid state vector and the optimal reactive power compensation. The computational complexity of solving  non-convex optimization problems is shifted to offline training of a deep neural network.  
% The computational complexity of the non-convex optimal reactive power control problem is reduced through parameterization with a learning model. Specifically, deep neural network is advocated here to learn the functional relationship between the power injection and the optimal reactive power compensation. 
In the training phase, by feeding grid state vectors obtained from  
historical data or through simulations, the weight parameters of the deep neural network are updated iteratively via policy gradient method. %is used here for efficiently estimating the gradients of power loss.
In the online inference phase, or real-time implementation,  
%	after several layers of simple operations, the output of the network is the optimal reactive power control for real-time power injection. 
one just needs to pass the observed state vector into the trained deep neural network, and obtains a near-optimal reactive power control at the output. Our model-free approach requires no system knowledge and is computationally inexpensive.	It also bypasses the need for data labels, and tackles the optimal reactive control problem through policy gradients.
% Furthermore, unlike existing works applying machine learning methods to reactive power control, our proposed approach is unsupervised and directly solve the resulting optimization problems bypassing the need for labeled data.

Regarding the remainder of this paper, Section \ref{sec:model} introduces our system model.
Section \ref{sec:prob} outlines the reactive power control problem formulation, followed by the proposed statistical learning solver in Section \ref{sec:solver}.
Numerical tests using a real-world feeder are presented in Section \ref{sec:test}, with concluding remarks drawn in Section \ref{sec:conc}. 

\emph{Notation.} Lower- (upper-) case boldface letters denote column vectors (matrices), with the exception of power flow vectors $(\pmb{P}, \pmb{Q})$, and normal letters represent scalars. Calligraphic symbols are reserved for sets, and $\triangle (\mathcal {S})$ represents the distribution over space $\mathcal S$.
%while $\bf 1$ denotes all-ones vector.
%% $\mathbb{R}^{\rm N}_+$ denotes the set of all non-negative $N$-dimensional vectors. 
%%the symbol $^{\top}$ stands for transposition; 
%%Prime stands for the transposition of vector and matrix.
%Symbol $^{\top}$ stands for transposition, and $\| \pmb x \|$ is the $l_2$-norm of~$\pmb x$.

%%%%%%%%%%%%%%%%%%%%%%%%%%%%%%%%%%%%%%%%%%%%%%%%%%%%%%%%%%%%%%%%%%%%%%%
\section{SYSTEM MODEL}\label{sec:model}

Consider a radial power distribution network 
%	with $N+1$ buses, 
%	labeled by $\{0\} \cup \mathcal{N}$, where $\mathcal{N}:=\{1,\ldots,N\}$. The network can be
%	 which can be 
modeled by a tree graph 
$\mathcal{G}:=(\mathcal{N}_0,\mathcal{E})$, 
where $\mathcal{N}_0:=\{0\} \cup \mathcal{N}$ denotes the set of buses, and 
$\mathcal{E}$ the set of edges.	
%	 $|\mathcal{E}| = N$ represents the carnality of the edge (line) set $\mathcal{E}$. 
The tree is rooted at the substation bus indexed by $n = 0$, and all branch buses are collected in $\mathcal{N}:=\{1,2,...,N\}$. For each bus $i\in\mathcal{N}$, let $v_{n}$ denote its squared voltage magnitude, and $p_n + j q_n$ denote its complex power injection, where $p_{n}:=p_{n}^g-p_{n}^c$ and $q_{n}:=q_{n}^g-q_{n}^c$
with superscript $^g$ ($^c$) specifying generation (consumption). 

\begin{figure}
	\centering
	\includegraphics[width =0.45 \textwidth]{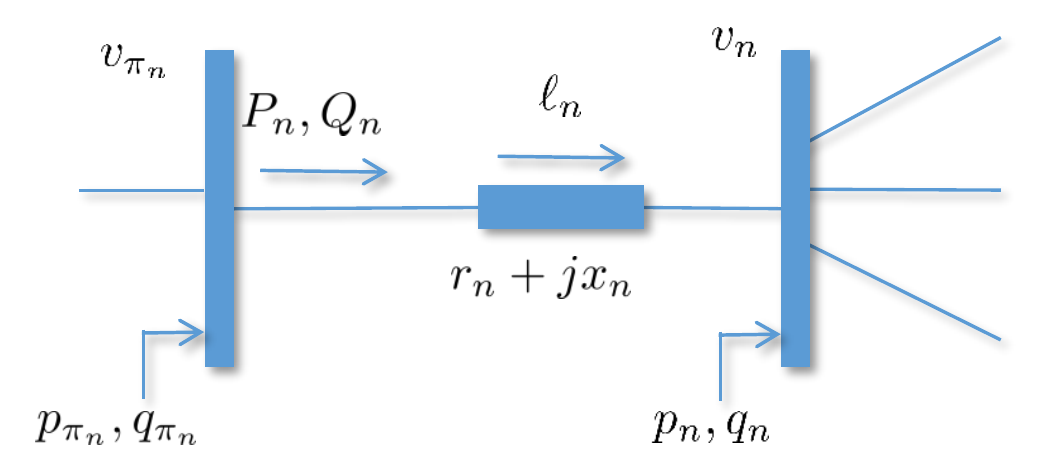}
	\caption{Bus $n$ is connected to its unique parent $\pi_n$ via line $n$.}
	\label{fig:lineardiagram}
\end{figure}

Thanks to the radial distribution grid topology, every non-root bus $n\in\mathcal{N}$ 
has a unique parent bus, denoted by $\pi_n$; and they are joined through the $n$-th distribution line $(\pi_n,n) \in\mathcal{E}$, whose impedance is given by $r_n+jx_n$. Let $P_n+jQ_n$ represent the complex power flow from buses $\pi_n$ to $n$ seen at the `front' end, and $\ell_n$ represent the magnitude square of the current over line $n\in \mathcal{E}$. For future reference, collect all nodal and line quantities into column vectors $\pmb{v}$, $\pmb{p}$, $\pmb{q}$, $\pmb{p}^g$, $\pmb{q}^g$, $\pmb{p}^c$, $\pmb{q}^c$, and $\pmb \ell$. 
See Fig.~\ref{fig:lineardiagram} for a depiction.

The radial grid can be described by the so-termed \emph{branch flow model} \cite{baran1989optimal}, 
which enforces the following equations for all $ n \in \mathcal N$  
\begin{subequations}\label{eq:nonlinear}
	\begin{align}
	P_n&=\sum_{j\in\chi_n}P_j  - p_n + r_n \ell_n \label{eq:nonp}\\
	Q_n&=\sum_{j\in\chi_n}Q_j  - q_n+x_n \ell_n\label{eq:nonq}\\
	v_n&={v_{\pi_n}}\!- 2(r_nP_n\!+x_nQ_n)\!+(r_n^2\!+x_n^2)\ell_n  \label{eq:nonv}\\
	\ell_n&=\frac{P^2_n+Q^2_n}{v_{\pi_n}}\label{eq:ml}
	\end{align}
\end{subequations}
% (but the index $t$ is ignored for brevity), 
where the set $\chi_n\subseteq\mathcal{N}$ collects all children buses for bus $n$.

Traditionally, for a smart inverter located at bus $n$ with nominal power capacity $\bar s_n$, and a solar panel equipped at this bus with an nameplate active power capacity $\bar p^g_n$, it should hold that $ \bar s_n = \bar p^g_n$.
In addition, the reactive power $q^g_n$ generated by the inverter is constrained by
$|q_{n}^{g}|\leq \sqrt{(\bar s_n)^2-( p^g_n)^2}$, where $p^g_n$ is the smart inverter output. However, to capture the special scenario that no reactive power can be provided when the maximum inverter output is reached (i.e., $p^g_n = \bar p^g_n$),  oversized inverters' nameplate capacity (i.e., $ \bar s_n > \bar p^g_n$) is used in practice. % \cite{turitsyn2011options}.
For instance, the reactive power compensation provided by inverter $n$ can be $|q_{n}^{g}|\leq 0.4 \bar p^g_n$, if choose $ \bar s_n = 1.08 \bar p^g_n$ and limit $q_{n}^{g}$ to $ \sqrt{(\bar s_n)^2-( \bar p^g_n)^2}$ instead of $\sqrt{(\bar s_n)^2-( p^g_n)^2}$, regardless of the instantaneous PV output $p^g_n$ \cite{kekatos2015stochastic}. 
Under this policy, the reactive injection region is the time-invariant convex set
\begin{equation}
{\underline{\pmb q}^g \leq \pmb q^g \leq \bar {\pmb q}^g}
\end{equation}
%As such, $q_{n}^{g}$ generated by inverter $n$ is constrained as 
%\begin{equation}
%\bar q^g_n:= \sqrt{(\bar s_n)^2-(\bar p^g_n)^2}, ~\forall n \in \mathcal N.
%\end{equation}
where $\pmb q^g \in \mathcal{R}^M$, and $M$ denotes the number of inverters in the grid. Moreover, the voltage magnitude at every bus $n \in \mathcal{N}$ should be maintained within a prespecified range, i.e., $v_n\in[\underline{v}_n,\overline{v}_n]$. In practice this range is chosen to be $\pm 5\%$ of its nominal value. For future use, rewrite voltage regulation constraints at all buses $n \in \mathcal N$ in a compact way as 
\begin{equation}
\pmb {v}\in\mathcal{V}:=\{\pmb {v}:\underline{\pmb {v}}\leq \pmb {v} \leq\overline{\pmb {v}}\}.
\end{equation}  

In distribution grids, it holds that $p_{n}^g = p_{n}^c = q_{n}^c = 0$ and $q_{n}^g >0$ when bus $n$ only has a capacitor; while $p_{n}^g = q_{n}^g =0$, $p^c_n \geq 0$, $q^c_n \geq 0$ when bus $n$ is a purely load bus; and a distributed generation bus $n$ not only consumes power denoted by $p_{n}^c $, $q_{n}^c$, but also generate active power  $p_{n}^g \geq 0$, and provide negative or positive reactive power $q^g_n$.
%	where $p_{i}:=p_{i}^g-p_{i}^c$ and $q_{i}:=q_{i}^g-q_{i}^c$
%	with superscript $g$ ($c$) denoting generation (consumption). in the front there is an error need to be fixed.
%	
%	\remark In distribution grids, it holds that $p_{n}^g = p_{n}^c = q_{n}^c = 0$ and $q_{n}^g >0$ when bus $n$ has a capacitor; while $p_{n}^g = q_{n}^g =0$ when bus $n$ is a purely load bus; and $p_{n}^c \geq 0$, $q_{n}^c \geq 0$, $p_{n}^g \geq 0$ when bus $n$ is equipped with a DG.
% 	
Moreover, active power consumption and solar generation $(\pmb{p}^c,\pmb{q}^c,\pmb{p}^g)$ can be predicted through the hourly and real-time market (see e.g., \cite{kekatos2015stochastic}), or by means of running load demand (solar generation) prediction algorithms \cite{tsp2019psse}.

%%%%%%%%%%%%%%%%%%%%%%%%%%%%%%%%%%%%%%%%%%%%%%%%%%%%%%%%%%%%%%%%%%%%%%%
\section{PROBLEM FORMULATION}\label{sec:prob}

%During time period $t$, reactive power compensation $\pmb q^g$ is either from utility-owned devices or smart inverters~\cite{yang2019real}. 
%{\color{red}control smart inverters k}
In the envisioned distribution network operation scenario, active power $\pmb p$ is controlled at a coarse timescale. Depending on the variability of active power and cyber resources (sensing, communication, and computation delays), reactive power compensation occurs over time intervals indexed by $t = 0, 1, \ldots$, which could either be real-time market periods, e.g., $5$ minutes, or even shorter, e.g., $30$ seconds.
%These intervals could either coincide with real-time market periods (e.g., 5 minutes), or be even shorter (30 seconds), depending on the variability of active powers and cyber resources (sensing, communication, and computation delays).
% of the reactive power control problem is to find feasible $\pmb q^g_t$ so that 
Let $(\pmb p_t, \pmb q_t)$ denote the active and reactive power injections at all non-root buses during control period $t$. The total power loss across all distribution lines can be expressed as $\sum_{n=1}^N r_{n,t} \ell_{n,t}$.
%\begin{equation}\label{eq:losses}
%f(\pmb{p}_t,\pmb{q}_t)=\sum_{n=1}^N r_n \ell_{n,t}.
%\end{equation}
Given load consumptions $(\pmb{p}^c_t,\pmb{q}^c_t)$ and generation $\pmb{p}^g_t$ at the beginning of each interval $t$, the goal of reactive power control is to find feasible reactive power injections $	{\pmb q}^{g,\ast}_t$ for smart inverters such that the power loss across all distribution lines is minimized while maintaining all bus voltage magnitudes within a prescribed range. Formally, the reactive power control problem is formulated as follows  
\begin{align} \label{eq:instantaneous}
{\pmb q}^{g,\ast}_t:&= \arg \min_{\underline{\pmb q}^g \leq \pmb q^g \leq \bar {\pmb q}^g} f(\pmb p_t,\pmb q^g- \pmb q^c_t)
\end{align}
where $f(\pmb p_t,\pmb q^g- \pmb q^c_t)$ admits the following form
\begin{subequations}\label{eq:instantaneous_f}
	\begin{align}
	f(\pmb p_t, \pmb q^g- \pmb q^c_t) 
	&= \min_{{ \pmb{P}_t,\pmb{Q}_t \atop \boldsymbol{\ell}_t, \pmb{v}_t}} \sum_{n=1}^L r_{n,t} \ell_{n,t}\label{eq:instantaneous_obj}\\  
	\textrm{s.to}~~P_{n,t}&=\sum_{j\in\mathcal{C}_{n,t}}P_{j,t}  - p_{n,t} +r_{n,t} \ell_{n,t},~n\in\mathcal{N}\label{eq:spf}\\
	Q_{n,t}&=\sum_{j\in\mathcal{C}_n}Q_{j,t} - q_{n,t} +x_{n,t} \ell_{n,t},\;~n\in\mathcal{N}\label{eq:sqf}\\
	v_{n,t}&=v_{\pi_{n,t}}  + (r_{n,t}^2+x_{n,t}^2)\ell_{n,t}  \nonumber \\ & \quad ~- 2(r_{n,t} P_{n,t} + x_{n,t} Q_{n,t}), ~~ \;n\in \mathcal{N}\label{eq:svf}\\
	\ell_{n,t}& = \frac{P_{n,t}^2+Q_{n,t}^2}{v_{\pi_{n,t}}}, \qquad \quad \quad ~~~~~n\in\mathcal{E}\label{eq:slf}\\
	%	&|q_{n}^{g}|\leq \bar q^g_n,\,\,\, \qquad \quad \quad \qquad \quad  ~ n \in \mathcal N\\
	\pmb{v}& \in \mathcal{V}\label{eq:sVf}.
	\end{align}
\end{subequations}
%Given load consumption $(\pmb{p}^c_t,\pmb{q}^c_t)$ and generation $\pmb{p}^g_t$ per interval $t$, the goal of reactive power control is to find feasible reactive power compensation $	{\pmb q}^{g,\ast}_t$ for smart inverters such that the power losses across distribution lines are minimized while the bus voltage magnitudes are maintained within the desired region. Formally, the reactive power control problem is formulated as follows  
%%Denote $f_t(\pmb s_t, \pmb q^g(\pmb s)):  = f(\pmb{p}_t,\pmb q^g- \pmb q^c_t):=f(\pmb{p}_t,\pmb{q}_t)$ for brevity.
%%The conventional reactive power control problem is to find
%\begin{subequations}\label{eq:instantaneous}
%	\begin{align}
%	\label{eq:instantaneous_obj}
%	{\pmb q}^{g,\ast}_t:= \arg \min_{\underline{\pmb q}^g \leq \pmb q^g_t \leq \bar {\pmb q}^g} ~& f(\pmb{p}_t,\pmb q^g- \pmb q^c_t)
%	\end{align}
%\end{subequations}
%{\color {red}where $\bar q^g_n$ denotes the maximum feasible reactive power compensation, and $\mathcal{V}$ is the prespecified range of voltage magnitudes defined earlier.  }
Clearly, constraints \eqref{eq:spf}--\eqref{eq:svf} and \eqref{eq:sVf} are linear with respect to system variables $(\pmb p_t,\pmb q_t,\pmb{P_t},\pmb{Q_t},\boldsymbol{\ell_t},\pmb{v_t})$. Nevertheless, constraints in \eqref{eq:slf} are quadratic equalities, depicting a non-convex feasible set and rendering the optimization problem non-convex and NP-hard in general \cite{gan2012exactness}. 

To address this issue, these equalities in \eqref{eq:slf} have been recently relaxed to convex inequalities described by the hyperbolic constraints \cite{gan2012exactness} 
\begin{align}\label{eq:inequality}
P^2_{n,t}+Q^2_{n,t}\le {v_{\pi_{n,t}}}\ell_{n,t}.
\end{align}
Substituting \eqref{eq:inequality} into \eqref{eq:instantaneous_f} yields
\begin{subequations}\label{eq:socp}
	\begin{align}
	f(\pmb{p}_t,\pmb{q}_t)~&=\min_{{\pmb{P}_t,\pmb{Q}_t \atop \boldsymbol{\ell}_t, \pmb{v}_t}} \sum_{n=1}^L r_{n,t} \ell_{n,t}\label{eq:scost}\\ 
	\textrm{s.to}~~&\eqref{eq:spf} - \eqref{eq:svf}, {\rm~and~} \eqref{eq:sVf}  \label{eq:relax1}\\ 
	\ell_{n,t} &\geq \frac{P_{n,t}^2+Q_{n,t}^2}{v_{\pi_{n,t}}},  ~n\in\mathcal{E}\label{eq:relax2}
	\end{align}
\end{subequations}
%\begin{subequations}\label{eq:socp}
%	\begin{align}
%	f(\pmb{p},\pmb{q})~&=\min_{{\pmb{P},\pmb{Q} \atop \boldsymbol{\ell}, \pmb{v}}} \sum_{n=1}^L r_n \ell_n\label{eq:scost}\\ 
%	\textrm{s.to}~P_n&=\sum_{j\in\mathcal{C}_n}P_j  - p_n +r_n \ell_n, \quad \quad \quad \quad \quad ~~n\in\mathcal{N}\label{eq:sp}\\
%	Q_n&=\sum_{j\in\mathcal{C}_n}Q_j  - q_n +x_n \ell_n, \quad \quad \quad \quad \quad ~n\in\mathcal{N}\label{eq:sq}\\
%	v_n&=v_{\pi_n}\!\!\!+(r_n^2\!+\!x_n^2)\ell_n\!\!-\! 2(r_n P_n\!\!+\!x_n Q_n),~ n\in \mathcal{N}\label{eq:sv}\\
%	\ell_n &\geq \frac{P_n^2+Q_n^2}{v_{\pi_n}}, \qquad \quad \quad \quad \quad \quad \quad \quad \;\;n\in\mathcal{E}\label{eq:sl}\\
%	\pmb{v}&\in \mathcal{V}\label{eq:sV}
%	\end{align}
%\end{subequations}
where \eqref{eq:relax2} can also be equivalently expressed as a second-order cone
\begin{align}\label{eq:soc}
\!\left\|\begin{array}{c}
2P_{n,t}\\
2Q_{n,t}\\
\ell_{n,t}-v_{\pi_{n,t}}\end{array}\right\|\le v_{\pi_{n,t}}+\ell_{n,t}.
\end{align}
%Equations \eqref{eq:nonp}-\eqref{eq:nonv} and \eqref{eq:soc} now define a convex feasible set. 
%%Recent efforts have leveraged this relaxed set (instead of the original non-convex one) to study several key grid management tasks; see e.g., \cite{gan2012exactnessconvex, molzahn2019survey} for recent surveys. 
%The procedure of leveraging this relaxed set (instead of the non-convex one) is known as SOCP relaxation \cite{gan2012exactnessconvex, molzahn2019survey}. Interestingly, 
%It should be notice that SOCP relaxation is exact in the sense that the set of inequalties \eqref{eq:soc} holds with equalities at the optimum \cite{gan2015exact}. 
Constraints \eqref{eq:relax1} and \eqref{eq:relax2} represent now a convex feasible set, %The relaxed problem \eqref{eq:socp} is now a convex SOCP, 
and the problem in \eqref{eq:socp} can be solved by standard convex programming methods.  
Interestingly, it has been shown that under certain conditions, at the optimal solution of \eqref{eq:socp}, equalities are attained in  \eqref{eq:soc}; see details in e.g.,  \cite{gan2012exactnessconvex}. In this case, the optimal solution of the original problem \eqref{eq:instantaneous_f} is recovered too. 
\begin{figure}
	\centering
	\includegraphics[width =0.5\textwidth]{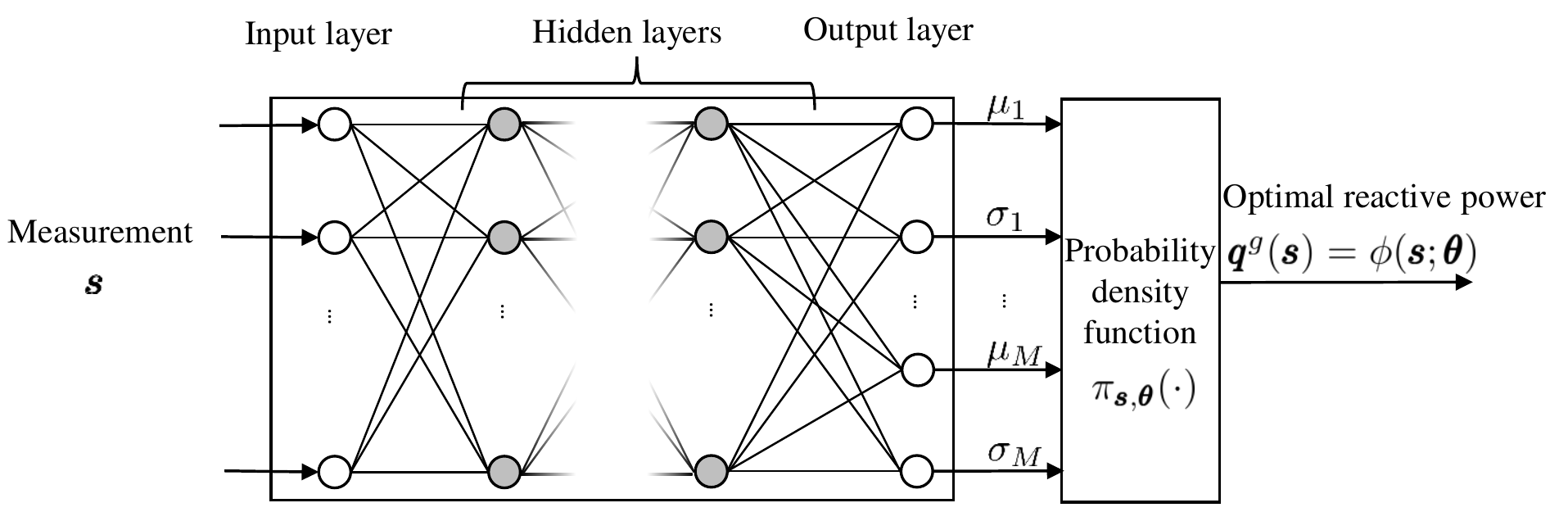}
	\caption{Statistical learning architecture.}
	\label{fig:deepneuralnetwork}
\end{figure}

It is worth pointing out that problem \eqref{eq:instantaneous} formally characterizes the optimal reactive power control policies for a diverse set of networked control problems, including e.g., voltage regulation, Volt/VAR control, and optimal power flow \cite{gan2012exactnessconvex}, by choosing suitable objective functions. 
If active and reactive power injections $(\pmb{p}_t,\pmb{q}_t^c)$ were both known precisely in advance and remained constant within period $t$, the optimal reactive power compensation $\pmb{q}^{g,\ast}_t$ would be  found by solving \eqref{eq:instantaneous}. 
However, such conditions are hardly met in contemporary distribution systems, due partly to i) time-varying active and reactive injections;
and, ii) noise-contaminated observations caused by direct measurements, delayed estimates, or inaccurate forecasts. To bypass these challenges, minimizing the averaged power loss over the power injections $(\pmb{p}_t,\pmb{q}_t^c)$ provides an alternative to the static reactive power control formulation in \eqref{eq:instantaneous}, given by
\begin{equation}\label{eq:prob}
\pmb  q^{g,\ast}_t:= \arg  \min_{\underline{\pmb q}^g \leq \pmb q^g \leq \bar {\pmb q}^g} ~\mathbb{E}\left[f(\pmb{p}_t,\pmb q^g- \pmb q^c_t)\right].
\end{equation}
For notational convenience, let us define the state vector $\pmb s_t : = (\pmb p_t, \pmb q^c_t)$, which is assumed to be a stationary random process, and rewrite the loss function $f(\pmb{p}_t,\pmb q^g- \pmb q^c_t)$ as $f(\pmb{q}^g;\pmb{s}_t)$. 
%The state $\pmb s_t$ is to be feed as input to our deep neural network.  and $\pmb q^g_t $ the output. 
Substituting this display into the original problem \eqref{eq:prob}, yields
\begin{equation}\label{eq:probmap}
\pmb  q^{g,\ast}(\pmb s_t):= \arg \min_{\underline{\pmb q}^g \leq \pmb q^g \leq \bar {\pmb q}^g} ~\mathbb{E}\left[f(\pmb{q}^g;\pmb{s}_t)\right].
\end{equation}
Rather than the unreliable and possibly obsolete instantaneous $\pmb q^{g,\ast}_t$ found through \eqref{eq:instantaneous}, problem \eqref{eq:probmap} is expected to yield smoother power control decisions. But, evaluating the expectation in \eqref{eq:probmap} is nearly impossible in practice, even if the probability density function of $\pmb{s}_t$ was known. Challenge also comes from the computational burden of dealing with the non-convex constraint \eqref{eq:slf}.  To approximate $\pmb q^{g,\ast}$ in a computationally efficient manner, a statistical learning approach is developed next.  
\section{STATISTICAL LEARNING}\label{sec:solver}
The rapid growth in renewable generation is displacing traditional forms of energy generation while increasing the need for controllable and flexible resources to balance fluctuations in load and generation. In this section, we introduce a novel parameterization form of the reactive power control problem, as well as a learning solver based on a deep neural network. 
%This is mainly motivated by the universal approximation theorem, which implies that a feed forward neural network containing finite number of neurons can well approximate any continuous function over a compact set. deep neural networks can efficiently cope with the `curse of dimensionality' by providing compact low-dimensional representations for high-dimensional data 

%In this section, we discuss the details of the deep neural network parametrization model and both the theoretical and practical implications within our statistical learning framework.

%the increasingly popular set of parameterizations known as deep neural network.  which are often observed in practice to exhibit strong performance in function approximation. In particular, we discuss the details of the deep neural network parametrization model and both the theoretical and practical implications within our constrained learning framework.

\subsection{Parameterization}

%our objective is to learning the mapping between measurements and optimal reactive power control
%Regard $\pmb q^g_t $ as a function of $(\pmb{p}_t,\pmb{q}_t^c)$.

Instead of solving \eqref{eq:probmap} exactly, consider a parametrization for the reactive power compensation as follows
\begin{align}
\pmb q^g = \phi(\pmb s; \pmb\theta)
\end{align}
where 
$\phi(\pmb s;\pmb \theta)$ is some function given by e.g., a deep neural network, and 
$\pmb \theta \in \mathbb{R}^d$ collects all unknown parameters. 
%For brevity, the time index $t$ is dropped whenever it is clear from the context.
%%%%%%%%%%%%%%%%%%%%%%%%%%%%%%%%%%%%%%%%%%%%%%%%%%%%%%%%%%
%
Building on this, finding the optimal reactive power control $\pmb q^{g,\ast}$ in \eqref{eq:probmap} boils down to finding the optimal parameter vector $\pmb \theta^\ast$, such that the expected loss is minimized; that is, %by solving the statistical learning problem
\begin{align}\label{eq:probparam}
\pmb \theta^\ast = \arg \min_{\pmb \theta} ~&\mathbb{E}\left[f(\phi(\pmb s; \pmb\theta); \pmb s)\right].
\end{align}
%
%In order to obtain the parameterization $\phi(\pmb s; \pmb\theta)$ in \eqref{eq:probparam}, we successively update $\pmb \theta_t$ as follows
%\begin{equation}
%\label{eq:SGD}
%\pmb \theta_{t+1} = \pmb \theta_t - \beta_t \nabla_{\pmb \theta}\mathbb{E}\left[f( \phi(\pmb s_t, \pmb \theta_t);\pmb s_t)\right] 
%\end{equation} 
%where $\beta_t>0$ is a preselected learning rate.

To find $\pmb \theta^\ast$, a natural approach is to apply gradient descent type algorithms.
 %until satisfying some notion of optimality, and hence finding a solution. 
To this aim, one needs to obtain the gradient of the objective function in \eqref{eq:probparam} with respect to $\pmb \theta$, i.e., $\nabla_{\pmb \theta}\mathbb{E}\left[f( \phi(\pmb s; \pmb \theta);\pmb s)\right]$. 
%and $\nabla_{\pmb \theta}\mathbb{E}\left[f(\phi(\pmb s_t, \pmb \theta_t);\pmb s_t)\right] $ denotes the (sub-)gradient of expectation of loss function $\mathbb{E}\left[f(\phi(\pmb s_t, \pmb \theta_t);\pmb s_t)\right]$ with respect to $\pmb \theta$. 
%  The above gradient-based updates provide a natural manner by which to search for the optimal reactive power control $\pmb  q^{g,\ast}(\pmb s_t)$.
In practice however, there is no analytic form of $f(\phi(\pmb s; \pmb\theta);\pmb s)$ as a function of $\phi(\pmb s; \pmb\theta)$ or $\pmb \theta$. In \eqref{eq:instantaneous_f}, for instance, the loss function $f$ depends only implicitly on $\pmb q^g$. Instead, we can observe the function value $f$ for any grid operating point $(\pmb{p},\pmb{q},\pmb{P},\pmb{Q},\boldsymbol{\ell},\pmb{v})$ [cf.~\eqref{eq:instantaneous_f}], which can be used to estimate the gradient. This motivates development of a model-free approach \cite{mark2019learning}.
% Furthermore, if one may accurately model the grid, the expectation of the loss function found through existing models of the network, also do not always capture the true physical performance in practice.
%As mentioned,  this knowledge is not available in modern grids due to lack of high-precision instrumentation and reliable communications. 
%Using function values to find gradient estimates, calls for model-free approaches, including policy gradient method, which efficiently estimates the gradient of a function with a policy \cite{sutton2000nips}.  
%These challenges motivate our model-free approach, which finds $\pmb \theta^\ast$ without directly having access to gradient of objective, i.e., $\nabla_{\pmb \theta}\mathbb{E}\left[f( \phi(\pmb s; \pmb\ theta);\pmb s)\right]$. To this aim, a \textit{policy gradient} based method is developed  in sequel.  
% %; iii) lack of explicit knowledge of system.
%due to reliable communications, 
%the system controller has only their noise-contaminated observations (direct measurements or delayed state estimates). loss packet, noise. state measurement contain perturbations, the result will not be exact
%lack of high-precision instrumentation and reliable communications; ii)  render computing and implementing the optimal control challenging in practice. 
Specifically, for a given set of iterates and reactive power realizations $\{\tilde{\pmb \theta}, \tilde{\pmb q^g}\}$, the corresponding loss function values $\tilde f(\tilde{\pmb q}^g_{\pmb \theta}; \pmb s)$ can be observed from the system. Using $\{\tilde{\pmb \theta}, \tilde{\pmb q^g}\}$ and $\tilde f(\tilde{\pmb q}^g_{\pmb \theta}; \pmb s)$, the parameter vector $\pmb \theta$ can be updated through the policy gradient method \cite{sutton2000nips},
%For instance, we may pass test state vectors through the lines to measure their losses.
%iterates and  $\{\tilde{\pmb \theta}, \tilde{\pmb q^g}\}$, the loss function value $\tilde f(\tilde{\pmb q}^g_{\pmb \theta};\pmb s)$ can be observed. For instance, we may pass test state vectors through the lines to measure their losses. In general, these observations are unbiased estimates of the true loss function values. 
%We can then iteratively update the parameters $\pmb \theta$ utilizing the so-called \textit{policy gradient} updates \cite{sutton2000nips}. 
which constructs a gradient estimate with only function observations.

A control policy here is a mapping from state vectors $\pmb s$ to reactive power control decisions (a.k.a. actions) $\pmb q^{g}$.
Consider first the stochastic control policy $\pi: \pmb s \rightarrow \pmb q^g$, specifying a conditional distribution of all possible decisions $\pmb q^g$ given the current state $\pmb s$.
Denoting the probability of taking action $\pmb q^g$ at state $\pmb s$ as  $\pi_{\pmb s, \pmb\theta}(\pmb q^g)$,
%where the 
%To leverage policy gradient method, consider first a stochastic policy described by a density function $\pi_{\pmb s, \pmb\theta}(\cdot)$. 
%regard any policy $\phi(\pmb s; \pmb\theta)$ as a stochastic policy drawn from a distribution with density function $\pi_{\pmb s, \pmb\theta}(\cdot)$. This can be considered to be the delta function, i.e., $\pi_{\pmb s, \pmb\theta}(\pmb x) = \delta(\pmb x - \phi(\pmb s; \pmb\theta))$, for a deterministic policy.
% $\nabla_{\pmb\theta}\mathbb{E}\left[f(\phi(\pmb s_t, \pmb \theta_t);\pmb s_t)\right] $. The policy gradient method targets modeling and optimizing the policy directly, which is usually modeled with a parameterized function 
%Using this density function, 
the gradient of $\mathbb{E}\left[f(\phi(\pmb s; \pmb\theta);\pmb s)\right]$ with respect to $\pmb \theta$ can be written as
\begin{subequations} \label{eq:policygradient1}
	\begin{align}
	&\nabla_{\pmb \theta}\mathbb{E}\left[f(\phi(\pmb s; \pmb \theta);\pmb s)\right] = \nabla_{\pmb \theta} \int_{\pmb s} f(\phi(\pmb s; \pmb \theta);\pmb s) \textrm{Pr} (\pmb s) d{\pmb s}\\
	& = \nabla_{\pmb \theta} \int_{\pmb s} \int_{\pmb q^g} f( \pmb q^g;\pmb s) \pi_{\pmb s, \pmb \theta}(\pmb q^g)	\textrm{Pr}(\pmb s)d{\pmb q^g} d{\pmb s}\\
	&= \int_{\pmb s} \int_{\pmb q^g} f( \pmb q^g;\pmb s) \frac{\nabla_{\pmb \theta} \pi_{\pmb s, \pmb \theta}(\pmb q^g)	}{ \pi_{\pmb s, \pmb \theta}(\pmb q^g)}  \pi_{\pmb s, \pmb \theta}(\pmb q^g) \textrm{Pr}(\pmb s)d{\pmb q^g} d{\pmb s}\\
	&=\mathbb E _{\pmb q^g, \pmb s}\left[f( \pmb q^g;\pmb s) \nabla_{\pmb \theta} \textrm{log}\pi_{\pmb s, \pmb \theta}(\pmb q^g) \right]	
	\end{align}
\end{subequations}
where $\textrm{Pr}(\pmb s)$ denotes the probability of state $\pmb s$, and $\pmb q^g$ is drawn from the distribution $\pi_{\pmb s,\pmb \theta}(\cdot)$. Here, the computation of $\nabla_{\pmb \theta}\mathbb{E}\left[f(\phi(\pmb s; \pmb \theta);\pmb s)\right]$ is translated to evaluating the expectation of function $f( \pmb q^g;\pmb s)$ multiplied by the gradient
of the policy distribution $\nabla_{\pmb \theta} \textrm{log}\pi_{\pmb s, \pmb \theta}(\pmb q^g)$. This is indeed useful when we have an analytic form for $\pi_{\pmb s,\pmb \theta}(\pmb q^g)$. In such case, we may further replace the expectation on the right-hand side \eqref{eq:policygradient1} with a sample mean. 
%, and hence directly approximate the left-hand side of \eqref{eq:policygradient1}.
Specifically, by using previous function observations, we obtain the following gradient estimate	
\begin{equation}\label{eq:policygradient2}
\widehat{\nabla_{\pmb \theta}\mathbb{E}}\left[f(\phi(\pmb s; \pmb \theta);\pmb s)\right] = \hat f(\hat{\pmb q}^g_{\pmb \theta};\pmb s) \nabla_{\pmb \theta} \textrm{log}\pi_{\pmb s,\pmb \theta}(\hat {\pmb q}^g_{\pmb \theta})
\end{equation}
where $\hat {\pmb q}^g_{\pmb \theta}$ is the injected reactive power into the distribution grid, drawn from the distribution $\pi_{\pmb s,\pmb \theta}(\cdot)$, and $\hat f(\hat{\pmb q}^g_{\pmb \theta};\pmb s)$ is the corresponding observed loss function value obtained by solving  \eqref{eq:instantaneous_f}.

% regard any policy $\phi(\pmb s; \pmb\theta)$ as a stochastic policy drawn from a distribution with density function $\pi_{\pmb s, \pmb\theta}(\cdot)$. This can be considered to be the delta function, i.e., $\pi_{\pmb s, \pmb\theta}(\pmb x) = \delta(\pmb x - \phi(\pmb s; \pmb\theta))$, for a deterministic policy.

Previously, it was assumed that the policy $\phi(\pmb s; \pmb \theta)$ is stochastic. In deterministic cases, where the distribution is a  delta function, i.e., $\pi_{\pmb s, \pmb\theta}(\pmb x) = \delta(\pmb x - \phi(\pmb s; \pmb\theta))$.
% $\pi_{\pmb s,\pmb \theta}(\cdot)$ is a delta function, the gradient $\nabla_{\pmb \theta} \textrm{log}\pi_{\pmb s,\pmb \theta}(\hat {\pmb q}^g_{\pmb \theta})$ in \eqref{eq:policygradient2} cannot be evaluated.  
%without knowledge of expectation loss function $\mathbb{E}\left[f(\phi(\pmb s; \pmb \theta);\pmb s)\right]$.
To evaluate $\nabla_{\pmb \theta} \textrm{log}\pi_{\pmb s,\pmb \theta}(\hat {\pmb q}^g_{\pmb \theta})$ in \eqref{eq:policygradient2}, one may approximate the delta function with a known density function centered around $\phi(\pmb s; \pmb \theta)$.  To capture the power constraint $ {\underline{\pmb q}^g \leq \pmb q^g_t \leq \bar {\pmb q}^g}$, a truncated Gaussian distribution with a fixed support on the domain $[\underline{\pmb q}^g, \bar {\pmb q}^g]$ is  considered in next subsection.

\subsection{Model-free learning}

To find the policy $\pi_{\pmb s,\pmb \theta}(\cdot)$, we restrict ourselves to the increasingly popular set of parameterizations, known as deep neural networks \cite{goodfellow2016deep}. Indeed, deep neural networks have recently demonstrated remarkable performance in numerous fields, including computer vision, speech recognition, and robotics. A deep neural network can effectively tackle the `curse of dimensionality' by extracting low-dimensional representation for high-dimensional data \cite{goodfellow2016deep}. % \cite{DNNsurvey2017}.

%In order to parameterize the policy $\phi(\pmb s; \pmb \theta)$, 
Consider a feed-forward deep neural network connected to a truncated Gaussian  probability density function $\pi_{\pmb s,\pmb \theta}(\cdot)$ block; see Fig. \ref{fig:deepneuralnetwork} for an illustration. It takes as input the state vector $\pmb s$, followed by $L$ fully connected hidden layers with ReLU activation functions.   
The output of the deep neural network is a set of mean and standard deviation pairs $\{\mu_m, \sigma_m \}_{m=1}^M$, each corresponding to $M$ truncated Gaussian distributions. 
By feeding the outputs of the deep neural network into the probability density function $\pi_{\pmb s,\pmb \theta}(\cdot)$ block,  the reactive power compensation vector $\pmb q^g$ is sampled from $\pi_{\pmb s, \pmb \theta}(\pmb q^g)$. Stacking all the weights of the deep neural network into the vector $\pmb\theta$, we have a function approximation to estimate the reactive power compensation $\pmb q^g (\pmb s) = \phi(\pmb s; \pmb\theta)$. 
%Here we consider a feed-forward neural network to estimate the mean $\pmb \mu$ and variance $\pmb \sigma$ for this truncated Gaussian distribution, see Fig.~\ref{fig:deepneuralnetwork}.
%The outputs are mean $\pmb \mu \in \mathcal{R}^m $  and variance $\pmb \sigma \in \mathcal{R}^m $, corresponding to the distribution with density function $\pi_{\pmb s, \pmb \theta}(\pmb q^g)$ of the reactive power compensation for the input state vector. 

Using the gradient estimate in \eqref{eq:policygradient2}, the weights $\pmb \theta$ can be successively updated as follows 
\begin{equation}
\label{eq:SGD}
\pmb \theta_{t+1} = \pmb \theta_t - \beta_t \widehat{\nabla_{\pmb \theta}\mathbb{E}}\left[f({\phi}(\pmb s_t;  \pmb \theta);\pmb s_t)\right]\Big|_{\pmb \theta = \pmb \theta_{t}}
\end{equation} 
where $\beta_t>0$ is a preselected learning rate. This update in \eqref{eq:SGD} is a model-free approach, since it does not require explicit knowledge about the actual form of the function $f(\cdot)$ or distribution of $\pmb s$.
% Clearly, this implies that the proposed method is model-free because gradients are estimated directly from measurements rather than model knowledge. 
Different from a traditional  supervised approach where requires a set of a given training labeled data \cite{tsp2019psse}, the developed method here is unsupervised; hence circumvents the need for labeled data and directly solves~\eqref{eq:probmap}.

\begin{figure}
	\centering
	\includegraphics[width =0.45 \textwidth]{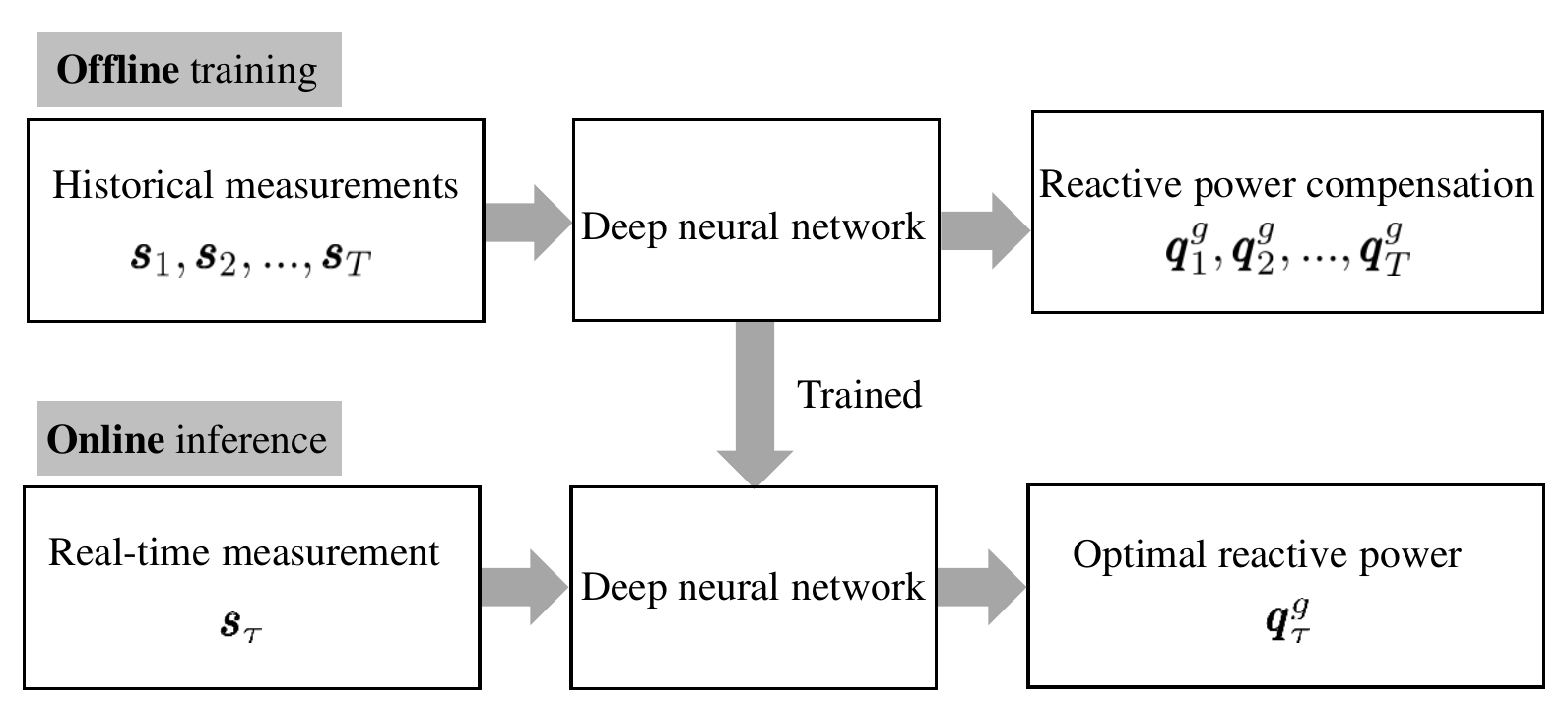}
	\caption{Two phases reactive power control procedure}
	\label{fig:twophases}
\end{figure}

\begin{algorithm}
	\caption{A statistical learning approach to reactive power control}
	\label{Alg_a}
	\hspace*{0.02in} {\bf Training phase:} 
	\begin{algorithmic}[1]
		\State \textbf{Initialize:} 
		$\pmb \theta$.
		\For{$t =1,2,\ldots, T$ } 
		\State Observe historical measurement $\pmb s_t$.
		\State Feed  $\pmb s_t$ into the deep neural network.
		\State Obtain deep neural network output mean $\pmb \mu_t $  and variance $\pmb \sigma_t $.
		\State Feed $\pmb \mu_t $ and $\pmb \sigma_t $ into  $\pi_{\pmb s_t, \pmb \theta_t}(\cdot)$. 
		\State Draw a sample $\hat {\pmb q}^g_{\pmb \theta_t}$ from the distribution $\pi_{\pmb s_t, \pmb \theta_t}(\pmb q^g)$.
		\State Obtain an estimate for $\hat f(\pmb s_t, \hat {\pmb q}^g_{\pmb \theta_t})$ via \eqref{eq:socp}.
		\State Calculate $\widehat{\nabla_{\pmb \theta}\mathbb{E}}\left[f(\pmb s_t,\phi(\pmb s_t; \pmb \theta_t))\right]$ via \eqref{eq:policygradient2}.
		\State Update $\pmb \theta_{t+1}$ according to \eqref{eq:SGD}.
		\EndFor
	\end{algorithmic}
	\hspace*{0.02in} {\bf Inference phase:} 
	\begin{algorithmic}[1]
		\For {$\tau = 1,2, \ldots$}
		\State Feed real-time measurement $\pmb s_\tau$ into the trained deep neural network.
		\State Obtain the deep neural network output mean $\pmb \mu_\tau $  and variance $\pmb \sigma_\tau $.
		\State Feed $\pmb \mu_\tau $ and $\pmb \sigma_\tau $ into $\pi_{\pmb s_\tau, \pmb \theta_\tau}(\cdot)$.
		\State Draw a sample $\hat {\pmb q}^g_{\pmb \theta_\tau}$ from the distribution $\pi_{\pmb s_\tau, \pmb \theta_\tau}(\pmb q^g)$.
		\EndFor
	\end{algorithmic}
\end{algorithm}

The proposed reactive power control procedure is tabulated in Alg. \ref{Alg_a}. It is implemented in two phases, namely offline training and online inference phases, as shown in Fig~\ref{fig:twophases}. Specifically, in the training phase, historical/simulated datum $\pmb s$ is fed into the deep neural network. For a given input datum $\pmb s_t$, our network spits out a reactive power compensation $\pmb q^g_t = \phi(\pmb s_t; \pmb \theta_t)$. Subsequently, the distribution network returns a loss for this state-action pair $(\pmb s_t,\pmb q^g_t)$ (which can also be found by solving \eqref{eq:instantaneous_f}). 
 Finally, a gradient estimate can be obtained using the policy gradient method in \eqref{eq:policygradient2}, based on which the neural network weight parameters are updated following \eqref{eq:SGD}. The trained deep neural network will be utilized in the inference phase. By taking the real-time state vector $\pmb s_t=(\pmb p_t, \pmb q^c_t)$ as input, the trained deep neural network outputs the optimal reactive power compensation $\pmb q^g_t $ to be implemented in the grid.  
Note that the proposed statistical learning approach is desirable for real-time reactive power control, as it shifts the computational burden of tackling non-convex optimization to offline training of a neural network. 

%The developed learning approach is summarized in Alg.~\ref{Alg_a}.

%The next section corroborates the performance of the developed method through numerical tests.

% The next section corroborates performance of developed method through numerical tests, but first a remark is in order.

%%%%%%%%%%%%%%%%%%%%%%%%%%%%%%%%%%%%%%%%%%%%%%%%%%%%%%%%%%%%%%%%%%%%%%% 
\section{NUMERICAL TESTS} \label{sec:test}
\begin{figure}
	\centering
	\includegraphics[width =0.45\textwidth]{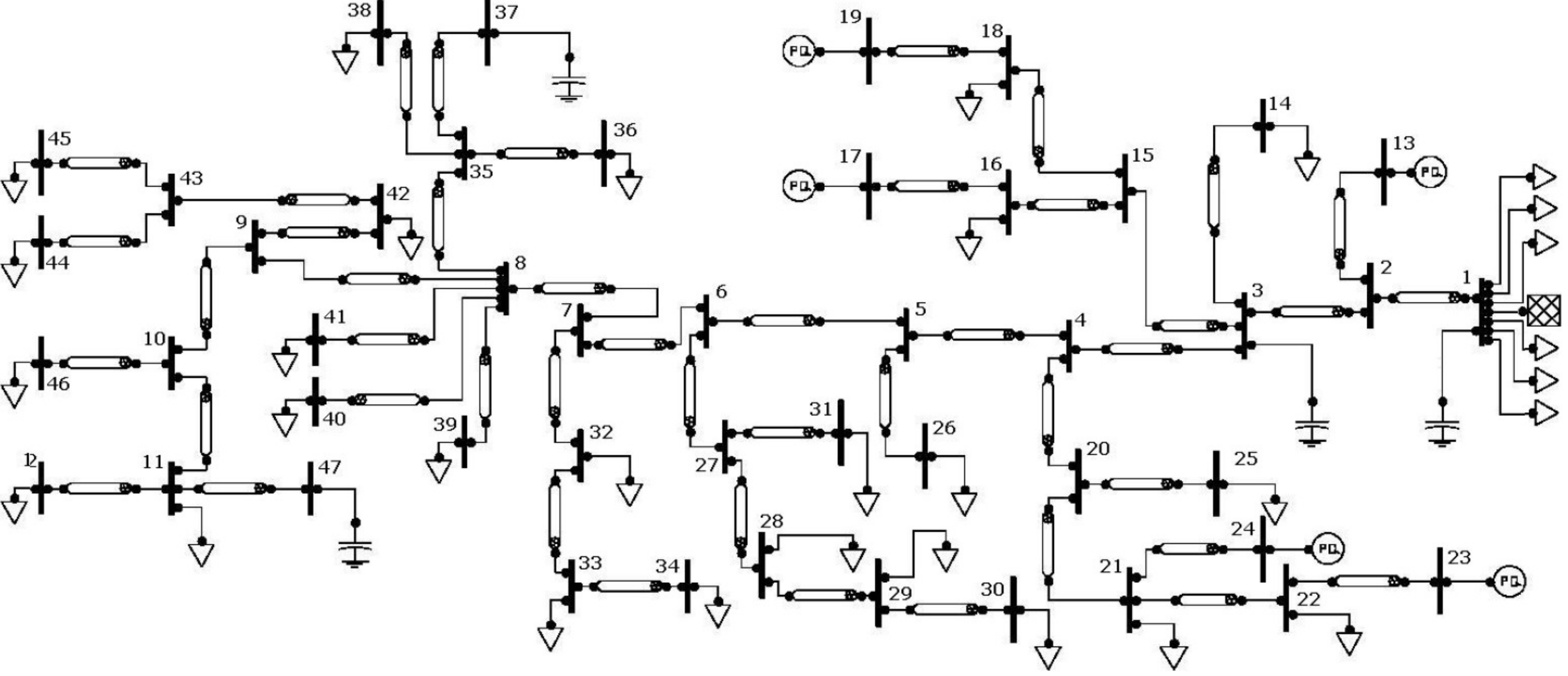}
	\caption{Schematic diagram of the $47$-bus distribution feeder.
		% Bus $1$ is the substation, and the $6$ loads connected to it model other feeders on this substation.
		%		 Buses $1, 3, 37$, and $47$ are equipped with shunt capacitors, while buses $2, 16, 18, 21$, and $22$ are equipped with inverters.
	}
	\label{fig:distributiongrid}
\end{figure}

In this section, the performance of our proposed statistical learning scheme was evaluated on a real-world $47$-bus feeder  with high penetration of renewables~\cite{farivar2011inverter};
%The novel scheme was tested on a $47$-bus feeder with high penetration of renewables~\cite{farivar2011inverter}; 
see Fig.~\ref{fig:distributiongrid}. 
This feeder is integrated with $M=5$ smart inverters located on buses $2$, $16$, $18$, $21$, and $22$, with capacities $300$, $80$, $300$, $400$, and $200$ kW, respectively.  A power factor of $0.8$ was assumed for all loads.

The training and test data were obtained by splitting the consumption and solar generation from the Smart${^*}$ project collected on August 24, 2011~\cite{barker2012smart}.
The CVX toolbox \cite{cvx} was used to solve the SOCP problem in \eqref{eq:socp} to evaluate $\hat f(\hat{\pmb q}^g_{\pmb \theta};\pmb s)$.
The deep neural network used here consists of three fully connected hidden layers, with $48$, $32$ and $16$ neurons per layer, respectively.
%The deep neural network was trained using ?TensorFlow? \cite{tensorflow2016} on an NVIDIA Titan X GPU with 12 GB RAM, with weights learned by the backpropagation based algorithm ?Adam? (with learning rate 10?3) for 40 epochs. The batch size was set to $30$.
%%To alleviate randomness in the obtained weights introduced by the training algorithms, deep neural network was trained and tested independently for 50 times, with reported results averaged over 50 runs.
To carry out the simulations, we used `TensorFlow' \cite{tensorflow2016} on an NVIDIA Titan X GPU with 12 GB RAM. The weight parameters of the deep neural network were updated using the back-propagation algorithm with `Adam' optimizer. The learning rate was fixed to $0.001$, and the batch size was $30$ throughout $40$ epochs of tests.  %The learning rate was $0.001$. 

%In the training phase, we took the reactive power control for a sample including $1,200$ data ($\pmb p$, $\pmb q^c$) as an example. 
%Fig. \ref{fig:loss} depicts the average power loss of this neural network for the whole data sets. Curve shows that the convergence of the neural network.

\begin{figure}
	\centering
	\includegraphics[width =0.5\textwidth]{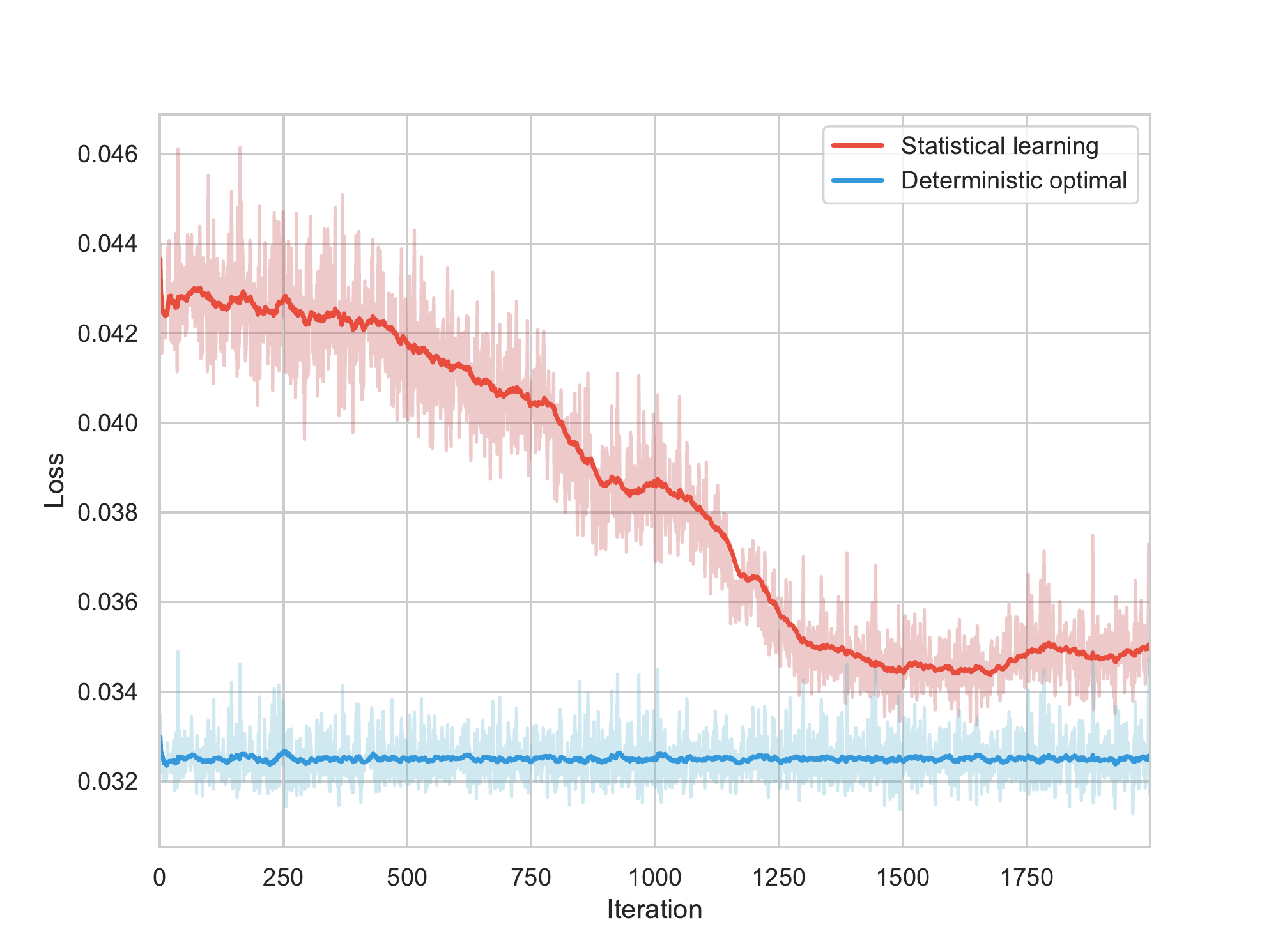}
	\caption{The training loss of the statistical learning approach compared with the baseline (optimal).}
	\label{fig:comparison}
\end{figure}

\begin{figure}
	\centering
	\includegraphics[width =0.5\textwidth]{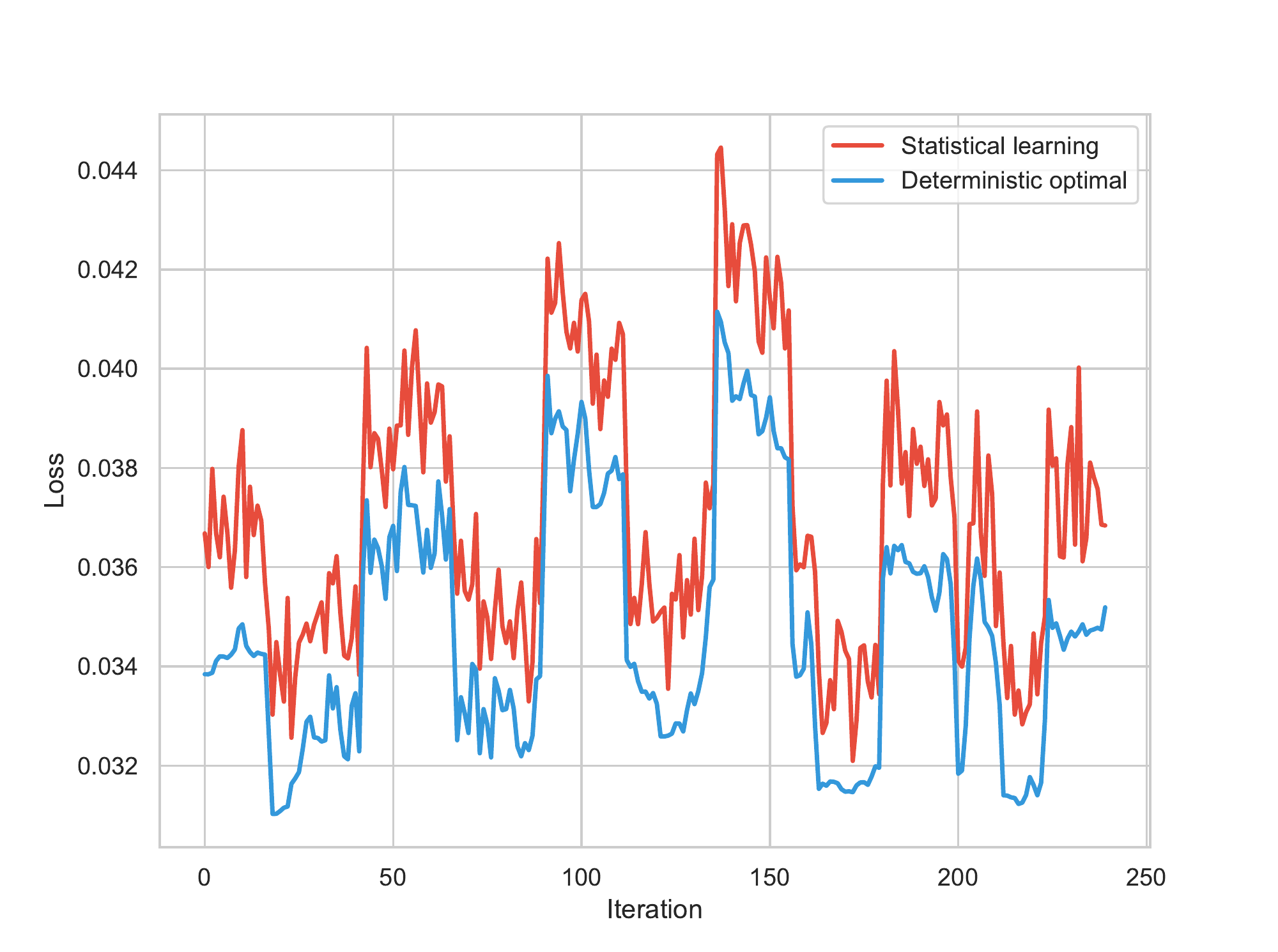}
	\caption{ The inference loss values of statistical learning approach and the baseline (optimal).}
	\label{fig:test}
\end{figure}

To assess the performance of the proposed approach, the following baseline was considered. 
%we use two baselinesTo benchmark the performance of our proposed scheme, the following loss $	f_{\pmb q^g}(\pmb{p},\pmb{q})$) was used here as a baseline
Assuming perfect observations of active and reactive power injections $(\pmb{p}_t,\pmb{q}_t^c)$ at the beginning of slot $t$, the optimal reactive power control can be found by 
solving the following 
problem
\begin{subequations}\label{eq:baseline}
	\begin{align}
	f(\pmb p_t,\pmb q^g_t-\pmb q^c_t)=\min_{\pmb q^g_t, \pmb P_t,\pmb Q_t \atop \boldsymbol{\ell}_t, \pmb{v}_t} ~&\sum_{n=1}^L r_{n,t} \ell_{n,t}\label{eq:scost2}\\ 
	\textrm{s.to}~~&\eqref{eq:relax1}-\eqref{eq:relax2}
	\end{align}
\end{subequations}
where $\pmb q^g_t$ is treated as an optimization variable. 
It should be noted that tackling this problem in real-time is computationally expensive, while the proposed approach finds $\pmb q^g_t$ after performing only several matrix-vector multiplications.
%It should be noted that the to solve the non-convex optimization problem in \eqref{eq:baseline}, it needs high computation complexity. 
The red curve in Fig. \ref{fig:comparison} shows the observed loss for the proposed approach, while the blue one depicts loss for the deterministic optimal one obtained via \eqref{eq:baseline} during the training phase. The light colour curves  correspond to the actual observed losses, while the dark ones are the running averaged ones. Clearly, our model-free approach learns to make optimal decisions $\pmb q^g_t$. 
%The convergence of the deep neural network weight parameter $\pmb \theta$ to that of the target network one $\pmb \theta^{\rm Tar}_\tau$ is depicted in Fig.~\ref{fig:error}. From Fig.~\ref{fig:error},  the novel scheme converges to a single policy and as will be shown next, it performs suitably in regulating the voltage.
In the inference phase,  the loss of the proposed approach versus the baseline is presented in Fig.~\ref{fig:test}. This plot demonstrates that the proposed model-free approach finds near-optimal reactive power control decisions. The running time of the proposed approach is one order of magnitude less than the optimization-based approach.
%Moreover, the proposed algorithm is desirable for real-time control as it shifts the computational burden to the training phase.

%It can be seen in Fig.~\ref{fig:test} that the difference is small.
%It should be noted that, since the proposed learning method alleviates almost all the computational burden at finding the optimal reactive power control by shifting it to the training time, the running time of the proposed approach is one orders of magnitude less than the optimization-based approach.

%\subsection{Historical Data Reactive Power Control}
%\subsection{Real-time Reactive Power Control}

%\begin{figure}
%	\centering
%	\includegraphics[width =0.49\textwidth]{figs/loss.png}
%	\caption{Average power loss.}
%	\label{fig:loss}
%\end{figure}

%%%%%%%%%%%%%%%%%%%%%%%%%%%%%%%%%%%%%%%%%%%%%%%%%%%%%%%%%%%%%%%%%%%%%%%
\section{CONCLUSIONS}
\label{sec:conc}
In this work, a statistical learning framework for reactive power control in distribution grids was developed. Uncertainties and delays in acquiring grid state motivate well this learning framework. The non-convexity of the underlying optimization, and lack of model knowledge makes reactive power control a challenge in modern grids, if not impossible, to solve directly. The theory of statistical learning empowered by non-linear functional approximation property of deep neural networks provided a fresh viewpoint to solve this problem. In particular, this work modeled the reactive power control policy via a deep neural network. The weights of the deep neural network were updated in an unsupervised and model-free fashion, circumventing the need for labeled data as well as an explicit model for the system. Our proposed method is computationally inexpensive, since all computational complexity is shifted to the training phase.  Preliminary numerical results on the real-world $47$-bus distribution network using real load data corroborate the merits of our developed approach. 
This work  opens up several interesting directions for future research. Robust methods for reactive power control in the presence of corrupted or adversarial observations is worth investigating. Exploiting topology of the power grid to design physics-informed architecture is also pertinent.  

\bibliographystyle{IEEEtranS}
\bibliography{power}

\end{document}